\documentstyle[aps,preprint,prd]{revtex}
\date{\today}
\begin{document}
\flushbottom

\widetext
\draft
\title{Correlated forward-backward dissociation and neutron spectra as a
luminosity monitor in heavy ion colliders}
\author{
Anthony J. Baltz, Chellis Chasman, and Sebastian N. White\footnotemark[1]
\footnotetext[1]{Corresponding author. Tel.: 1-516-344-5488;
fax: 1-516-344-3253;
e-mail: white1@bnl.gov.}}
\address{
Brookhaven National Laboratory,
Upton, New York 11973}
\date{January 12, 1998}
\maketitle

\def\thepage{\arabic{page}}
\makeatletter
\global\@specialpagefalse
\ifnum\c@page=1
\def\@oddhead{Draft\hfill To be submitted to Phys. Rev. C}
\else
\def\@oddhead{\hfill}
\fi
\let\@evenhead\@oddhead
\def\@oddfoot{\reset@font\rm\hfill \thepage \hfill}
\let\@evenfoot\@oddfoot
\makeatother

\begin{abstract} Detection in zero degree calorimeters of the correlated
forward-backward
Coulomb or nuclear dissociation of two colliding nuclei is presented as a
practical luminosity monitor in heavy ion colliders.  Complementary predictions
are given for total correlated Coulomb plus nuclear dissociation and for
correlated forward-backward single neutrons from the giant dipole peak.
\\
{\bf PACS: {25.75.-q; 29.40.Vj; 29.27.-a}}\\
{\it keywords}: Heavy ions; Calorimeter; Luminosity; Collider;
Dissociation
\end{abstract}

\makeatletter
\global\@specialpagefalse
\def\@oddhead{\hfill}
\let\@evenhead\@oddhead
\makeatother
\nopagebreak
\section{Introduction}

In a heavy ion collider such as RHIC the cross section for Coulomb dissociation
of one of the ions in a 100 GeV $\times$ 100 GeV Au + Au collision will be many
times the geometric cross section, about 95 barns\cite{brw}.  With such a
large cross section available, it has been suggested that the correlated
forward-backward Coulomb dissociation would provide a clean monitor of beam
luminosity with the very forward (zero degree) calorimeters proposed for
RHIC\cite{bw}.  In fact the calculated cross section for mutual Coulomb
dissociation was found to be quite large, about 4 barns.  But there is an
intrinsic limitation in the precision of the calculated mutual Coulomb
dissociation: a lower cutoff in impact parameter must be chosen,
and the calculated mutual Coulomb dissociation
varies approximately as the inverse square of this cutoff.  However, it was
found that if one adds nuclear dissociation to the
mutual Coulomb dissociation calculation, the cutoff merges
into the nuclear surface and the calculated cross section of 11 barns is
relatively insenstive to parameter variation\cite{bw1}.  Details of these
calculations are presented in Section III below.

The proposed zero degree calorimeter\cite{zero} detects the total
neutral energy in the forward direction.  For neutrons of the beam momentum
the detected signal varies then as the number of neutrons.  While the
dissociation cross section includes excitations of the nucleus up to many GeV,
the largest contribution is for excitations of less than about 25 MeV, the
region of the giant dipole resonance, amounting to about 65 barns.
Photonuclear data\cite{ves} indicate that the excited state decays
predominantly by single neutron emission in this region;  we will
show that the Coulomb dissociation to a single neutron final state is large
(50 barns) and contained in a single energy peak within about 10\% of the
beam energy.  In Section IV we present details and consider the detection of
correlated single neutron peaks.

Based on the above we will present several complementary methods of using
calculations of mutual Coulomb plus nuclear dissociation along with zero
degree calorimeter measurements as an absolute luminosity calibration at heavy
ion colliders such
as RHIC and LHC.  But first we will begin with some experimental
considerations.

\section{Experimental Considerations}
The primary
quantities of interest are: (1) the luminosity, (2) the distribution and
centroid of the interaction region along the beam direction, and (3) the
distribution in time of the interactions relative to the nominal bunch
crossing time.   A basic requirement on the reaction chosen for these
measurements (particularly of luminosity) is that it be relatively background
free and that it
have a straightforward acceptance, implying simple correction of measured rates
for experimental details.  Finally, if this reaction is to be used as an
absolute luminosity
determination, it will be necessary to reduce uncertainties in the calculated
cross section to $\leq 5 \% $ or better.  We deal below with the
issues pertaining to luminosity.  The other principal measurements rely on the
timing resolution of the neutron detectors.

What is the rate from which lumnosity is calculated?
We propose to measure the coincidence rate requiring forward and backward
energy of $\geq 60$ GeV within 2 mrad of each beam direction. These
requirements
are largely chosen to minimize uncertainties in the calculated cross section.
The coincidence requirement is also necessary to reduce contamination from
background sources.

The 60 GeV threshold choice (assuming 100 GeV/u beam energy) is simply a
requirement of at least 1 fragmentation neutron emitted into the detector and
allows for linewidth and experimental resolution.
(A resolution of $\leq 20\%$ at 100 GeV is expected for a practical detector).
Since at a collider the forward direction is only accessible downstream of beam
steering magnets, the contribution from protons and charged
fragments is expected to be negligible\cite{RHIC,Alice}.  Further specifics are
detailed in the following.

\subsection{Energy Threshold}
In order to include a real energy threshold in the coincidence requirement we
consider it essential that some fraction of triggers
populate an energy peak which can be clearly resolved from the continuum. The
peak will be used to adjust trigger energy scale and to 
compute inefficiencies after the fact. 
We examine below the energy spectrum of neutrons emitted in Coulomb
dissociation. Single neutron emission turns out to be prominent because of the
importance of the giant dipole resonance; the natural energy spread is smaller
than the expected detector resolution. 

\subsection{Experimental Backgrounds}
There are two principal types of experimental background that concern
us, those from beam interactions (which are therefore proportional to
luminosity) and those
which arise from single beam interactions with residual gas in the machine, 
for example, (and do not depend directly on luminosity).  The coincidence
requirement is designed to largely eliminate the latter.
   
Single beam rates can be estimated from the nominal conditions and vacuum 
requirements called for in the experimental areas.  Beam gas rates
are expected to be lower than the minimum bias
interaction rate at full luminosity (i. e. the signal to be measured,
roughly 7 barns cross section for Au on Au).  Since
the actual minimum bias rate can be significantly lower than the nominal design
value during commissioning the beam gas 
rate could become significant.  Similarly, rates due to interactions of the
beam halo, which are difficult to calculate a priori, could also be a large
background source.

The requirement of forward-backward coincidences greatly reduces the full
counting rate due to these sources except for accidentals.  It also makes
possible the secondary measurements described above.  The
spatial distribution of interactions is calculated from forward-backward time
differences.  Assuming a time resolution of order 200 psec, we calculate
a spatial resolution of a few cms.
Backgrounds due to accidental coincidences between the two beam directions,
including contributions from beam-beam interactions such as single beam Coulomb
dissociation (95 barns at RHIC), are found to be negligible.

\subsection{Luminosity Calibration and Cross section Uncertainties}
As was indicated in the Introduction, mutual Coulomb dissociation calculations
are probably limited to an uncertainty of not less than about 5\%.
There is a lower impact parameter cutoff in the Weizsacker-Williams
formalism that has no clear experimental counterpart.  In
fact, it is difficult to define an experimental cut which would clearly
distinguish between strong interaction final states and those that arise from
the high energy component of the Weizsacker-Williams spectrum.
        
For this reason in Section III we examine the possibility of calculating a
cross section
which is actually the sum of Coulomb and nuclear processes leading to
the forward-backward coincidence of neutral beam energy clusters.  We find
that this almost completely eliminates the above uncertainties.

\subsection{Beam Fragmentation in Nuclear Collisions}
Before we perform this calculation of Coulomb plus nuclear dissociation
we must address the question of
the fraction of nuclear interactions, as a function of impact parameter, in
which heavy ion collisions lead to a neutral beam energy cluster within 2 mr of
the beam direction. 
Fortunately this measurement has been recently performed in a dedicated test
at the CERN SPS \cite{NA49}.
For our purposes the probability of producing less
than one neutral forward cluster, even in a central collision, is negligible.
In what follows, we
take the probability of a neutral beam energy cluster within 2 mr of
the beam direction to be unity.

\section{Mutual Coulomb plus Nuclear Dissociation}
The cross section for heavy-ion dissociation may be accurately expressed
in terms of (generally experimentally known) photo-dissociation cross
sections $\sigma_{ph} (\omega)$ of the same nucleus over an appropriate energy
range of photon energies $\omega$.  This is the so-called Weizsacker-Williams
expression
\begin{equation}
\sigma_{dis} = {2 \alpha Z_p^2 \over \pi \gamma^2 }
\int {d \omega \omega }\sigma_{ph} (\omega) 
\int_{b_0}^\infty b\, d b
K_1^2({b\,\omega \over  \gamma}),
\end{equation}
where $K_1$ is the usual modified Bessel function.  The lower cutoff of the
impact parameter integral $b_0$ is normally fixed at a value to separate pure
Coulomb excitation from nuclear processes. 
$b_0$ provides a somewhat arbitrary parameter dependence which we deal
with below.  

We may define a probability of dissociation, $P(b)$ as a function of impact
parameter $b$
\begin{equation}
\sigma_{dis} = 2 \pi \int_{b_0}^\infty P(b) b\ d b.
\end{equation}
Then inverting the order of integration in Eq. (1) we have
\begin{equation}
P(b) = {\alpha Z_p^2 \over \pi^2 \gamma^2 }
\int {d \omega \omega }\sigma_{ph} (\omega) 
K_1^2({b\,\omega \over  \gamma}).
\end{equation}

If we assume independence of the various modes of dissociation, then the
probability of at least one dissociation excitation of the nucleus is given
by the usual Poisson distribution.  If the first order probability of Coulomb
dissociation at a given
inpact parameter $b$ is $P_C(b)$, then the probability of at least one
mutual Coulomb dissociation $P^m_C(b)$ is given by
\begin{equation}
P^m_C(b) = (1 - e^{-P_C(b)})^2 .
\end{equation}
The corresponding probability of at least one mutual nuclear dissociation
$P^m_N(b)$ is given in terms of a first order probability $P_N(b)$ by
\begin{equation}
P^m_N(b) = (1 - e^{-P_N(b)}).
\end{equation}
Note that there is no square here; all nuclear dissociation is mutual.
$P_N(b)$ is evaluated from the Glauber model:
\begin{equation}
P_N(b)=\int d x d y d z_p \rho_p (x - b, y, z_p) (1 - e^{\sigma_{NN}\int dz_t
\rho_t (x, y, z_t)}).
\end{equation}
The densities $\rho_p, \rho_t$ are parameterized with a Fermi function.
For Au the
charge distribution parameters are $R_C = 6.38, a = .535$ fm\cite{dev}.
From Hartree-Fock calculations one finds that the neutron radius should be
a little larger\cite{gal}.  We take the neutron radius of Au to be
6.6 fm and then average for a best overall radius of 6.5.  At RHIC energies
the total nucleon-nucleon cross section $\sigma_{NN}$ is 50 mb\cite{pdata}.

Consider now the probability of survival without mutual Coulomb or nuclear
dissociation $P_S(b)$.  It is the product of the separate probabilities.
Since all
nuclear dissociation is mutual, the nuclear-Coulomb mutual dissociation
contribution is redundant.  We have
\begin{eqnarray}
P_S(b)&=&(1 - P^m_C(b))(1 - P^m_N(b))
\nonumber \\
&=&e^{-P_N(b)} (2 e^{-P_C(b)} - e^{-2 P_C(b)}).
\end{eqnarray}
The probability of mutual excitation is then
\begin{equation}
P(b)=1 - e^{-P_N(b)} (2 e^{-P_C(b)} - e^{-2 P_C(b)}).
\end{equation}

Let us take the standard design case of 100 GeV + 100 Gev Au + Au at RHIC.
Seen in the frame of the nucleus being dissociated, the equivalent $\gamma$ of
the other ion providing the equivalent photons is 23,000. 
The experimental photo-dissociation cross section $\sigma_{ph}(\omega)$ 
that we utilized is shown in Fig. 1, which we have taken from Ref. \cite{brw}.

In Table I we present calculated first order and unitarity corrected mutual
Coulomb dissociation cross sections for two values of
$b_0^2$, 2.25 barns and 3
barns, corresponding to $r_0$ values of 1.29 and 1.49 respectively when one
expresses $b_0$ in term of $2 \times r_0$ A$^{1/3}$.  Cross sections are
tabulated as a function of a cutoff in
$\omega$, which should correspond in some fashion to an
experimental acceptance.  For comparison, the corresponding cross sections for
2.76 TeV + 2.76 TeV Pb + Pb collisions at LHC are shown in Table II.

Fig. 2 shows the calculated probabilities of correlated forward-backward
dissociation as a function of impact parameter.  At impact parameters of
13 fm and below the mutual dissociation probability becomes unity from the
nuclear dissociation alone.  Table III shows the
parameter dependence of correlated dissociation cross sections.  The
first four rows show how the
calculated mutual dissociation cross section becomes independent of the
cutoff parameter for values less than about 15 fm.  The fifth row compared
to the second row shows how a small increase in the nuclear size parameter
causes a small increase in the computed cross section.  The sixth row shows
how a 20\% reduction in the nucleon-nucleon cross section leads to a less than
1\% reduction in the mutual dissociation cross section.  The overall
conclusion is that we can predict a mutual Coulomb plus nuclear dissociation
cross section of 11 barns with an error of less than about 5\%.

\section{Forward-Backward Energy Peak}
From Eq.(1) we can write down the differential Weizsacker-Williams
cross section for heavy-ion dissociation 
\begin{equation}
{d \sigma_{dis} \over d \omega }= {2 \alpha Z_p^2 \omega \over \pi \gamma^2 }
\sigma_{ph} (\omega) 
\int_{b_0}^\infty b\, d b
K_1^2({b\,\omega \over  \gamma}).
\end{equation}
$\sigma_{ph} (\omega)$ is the photon cross section; in the present calculation
we take it from the photo-dissociation data leading to only a single neutron
in the final state\cite{ves}.  The square of the Bessel function gives the
the energy and impact parameter dependent number of equivalent photons from
the heavy ion.

The impact parameter integral may be approximated very accurately 
for $b_0\omega << \gamma$  to yield
\begin{equation}
\int_{b_0}^\infty b\, d b
K_1^2({b\,\omega \over  \gamma}) =
{ \gamma^2 \over \omega^2 } [ ln ({2\, \gamma \over b_0\omega })
-\gamma_{euler} -.5] = { \gamma^2 \over \omega^2 }
ln ({.681\, \gamma \over b_0\omega }).
\end{equation}
Putting in the factor of $\hbar c$ explicitly we obtain the familiar form
\begin{equation}
{d \sigma_{dis} \over d \omega}= {2 \alpha Z_p^2 \over \pi \omega}
\sigma_{ph} (\omega) 
ln ({.681\, \hbar c\, \gamma \over b_0\,\omega }).
\end{equation}

The energy dependence of the ($\gamma$, 1n) cross section on Au\cite{ves},
$\sigma_{ph} (\omega)$, is shown as the dashed line in Fig. 3.  The Au + Au
cross section for single neutrons in either forward beam direction,
obtained from Eq. (11), is shown as the solid line in
Fig. 3.  Notes the difference in scale on the plot of the two cross sections.
Setting the impact parameter cut off $b_0$ to 15 fm and integrating from the
threshold at 8.1 MeV up
to 24.0 MeV, we find that the Coulomb excitation to a single neutron
final state, $\sigma_{dis}$, is 50.6 barns.

The angular distribution of neutrons emitted from the giant dipole resonance
has been parameterized in the form $A + B \sin^2(\theta)$, and for Au
$A/B$ has been measured to be $.58/.38$\cite{tg}.  If one relativistically
boosts the emitted soft neutron energy to the lab frame, then one obtains the
spectra of Fig. 4.  The solid line is the predicted spectrum from the
experimental ratio of $A/B$.
For comparison, a purely isotropic distribution (dot-dashed line) and a
$\sin^2(\theta)$ distribution (dashed line) are also shown. 

From Fig. 4 it is clear that no matter what the the angular distribution of
the emitted
neutrons is, there will be a huge one neutron peak of about 50 barns.
Because of its size, this peak  will stand out with very little background:
other single neutron contributions from higher energy dissociation will be
relatively negligible\cite{lep}\cite{car}; soft two neutron spectrometer
contributions will be centered at about twice the single neutron energy of
100 GeV; higher numbers of neutral particles will be correspondingly higher
in energy.  Thus the large, well characterized single neutron
peak provides an ideal component for a Au + Au beam luminosity monitor in the
forward calorimeter at RHIC.

Now let us consider mutual Coulomb dissociation in which one or both of the
reaction products is a single neutron at the beam momentum.
The previous analysis made use of the lowest order expression Eq. (11) for the
single neutron cross section which is not properly
unitarized and contains a logarithmic dependence on $b_0$.  
A more accurate expression may be obtained in analogy with Section III by
again assuming a Poisson distribution in the number of excitations.  Then the
probability of one and
only one neutron excitation $P_{1n}(b)$ may be expressed terms of the first
order probability of one neutron excitation
$P^1_{1n}(b)$ and the normalization factor $e^{-P_N(b)-P_C(b)}$ which involves
all excitations 
\begin{equation}
P_{1n}(b) = P^1_{1n}(b)e^{-P_N(b)-P_C(b)}.
\end{equation}
Note that the normalization factor provides a natural impact parameter cutoff;
we will not need a dependence on $b_0$ in our cross section expression Eq. (2).
Multiple neutron final states completely dominate for impact parameters
smaller than grazing.

The expression for mutual Coulomb dissociation in which both of the neutral
reaction products are single neutrons is
the square of the Eq. (12) expression.  The analagous expression for
mutual Coulomb dissociation in which one 
reaction products is a single neutron and the other is any excitation is
\begin{equation}
P_{1n,xn}(b) = P^1_{1n}(b)e^{-P_N(b)-P_C(b)}(1 - e^{-P_N(b)-P_C(b)}).
\end{equation}

Table IV shows the dependence of computed cross sections on the radius
parameter of the nuclear density.  For Au + Au 6.38 corresponds to the the
radius of
the proton density while 6.50 is probably a more realistic value, an average
between a proton density and an expected neutron density.  In any case the
dependences are not large.  The mutual cross sections vary by about 1\%
over this range while the uncorrelated cross sections vary by less than
a tenth of one per cent.

The correlated single neutron cross section is predicted to be .45 barns
and the correlated cross section for a single neutron in one specified
detector along with any neutral in the other (including possibly a single
neutron) is three times that, 1.35 barns.  For completeness the 11 barn
cross section of Table III for any mutual excitation is repeated.

Corresponding cross sections for Pb + Pb at LHC have also been computed
and are also displayed in Table IV.  For Pb the
charge distribution parameters are $R_C = 6.624, a = .549$ fm\cite{dev} and
a more realistic nucleon matter radius is taken to be 6.65.  At LHC energies
the total nucleon-nucleon cross section $\sigma_{NN}$ is taken to be
85 mb\cite{pdata}. Again radius dependences are not large.

One might consider using the predicted 49 barn single neutron peak as a
luminosity monitor.  The limitation of this method comes from the luminosity
independent backgrounds due to beam gas and beam halo interactions discussed
in Section II.  The 102 barn nuclear plus Coulomb dissociation cross section
suffers from the same limitations without the positive energy signal of the
neutron peak.  Requiring a forward-backward single coincidence should
largely eliminate these backgrounds with the experimental advantage of
having at least one the neutron peaks.  

\section{Z Dependence}
We have performed calculations involving mutual Coulomb dissociation for
Au + Au and Pb + Pb reactions to exploit multi-barn cross sections.
Unfortunately these large cross sections do not persist to collisions of
lighter ion species: mutual dissociation cross sections scale approximately
as $Z^6$.  This is because single Coulomb dissociation scales as $Z^2$ times
the number of nucleons for the highest excitations or as $Z^2$ times
$(NZ)/A$ for the giant dipole resonance.  Table V presents some representative
comparisons.  Clearly for ions as light as Ca + Ca and O + O correlated
forward-backward energy deposit will be dominated by purely nuclear collisions,
and will lead to a simple geometric mutual cross section.

\section{Discussion}
We have presented several complementary solutions to using
computed Coulomb dissociation as a luminosity monitor at RHIC (and LHC).
As we showed in Section III, if we measure the combined Coulomb plus nuclear
mutual dissociation cross section, then we eliminate the problem of the lower
cutoff in the calculation of the mutual Coulomb dissociation cross section.
The mutual Coulomb plus nuclear cross section is predicted to be 11 barns.
This absolute
theoretical prediction should be good to about 5\% or better, but one must
detect a signal for all mutual events regardless of energy.  As we showed
in Section IV, if we measure
the neutron spectra, then we should find a clean peak from the giant dipole
excitation in both directions.  Requiring forward-backward coincidences within
the one neutron peak at the beam energy gives a much smaller .45 barn cross
section, but with the positive experimental energy identification of the
beam momentum peaks.  The other predicted cross sections in Table IV could 
provide complementary roles as a luminosity monitor, especially by exploiting
their predicted ratios.


\begin{figure}
\caption[Figure 1]{Photonuclear dissociation cross sections utilized for the
Au + Au Coulomb dissociation calculations.}
\label{pho}
\end{figure} 
\begin{figure}
\caption[Figure 2]{Calculated mutual dissociation probabilities for
Au + Au at RHIC.  }
\label{mut}
\end{figure} 
\begin{figure}
\caption[Figure 3]{The dashed line is the ($\gamma$,1n) cross section on Au
taken from Ref.\cite{ves} up to 20 MeV and extrapolated to zero at 24 MeV;
the units are .1 barns per MeV (the peak is at .52 barns).  The solid line
is the single neutron cross section for Au + Au at RHIC (in barns per MeV).}
\label{gia}
\end{figure}
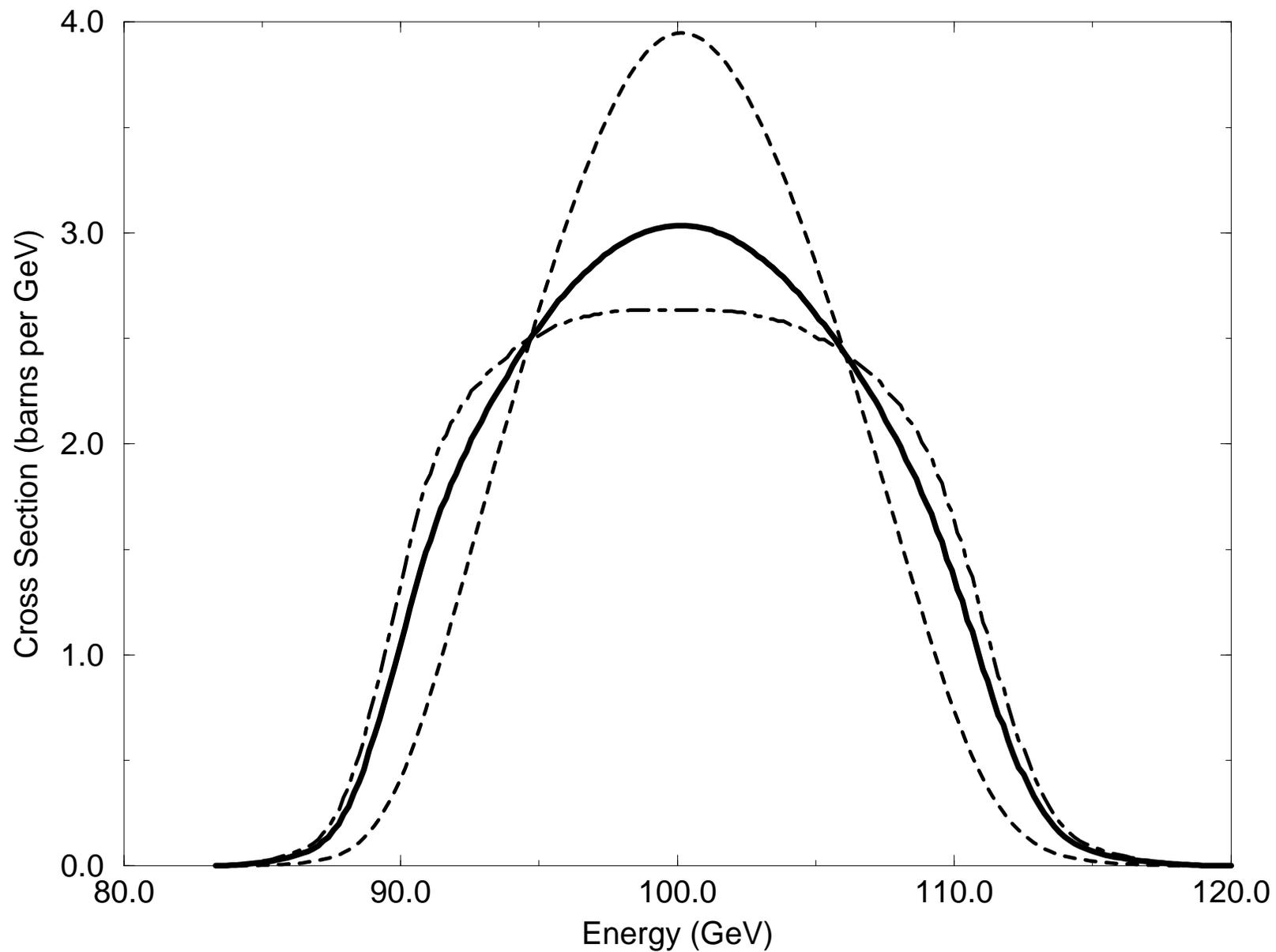
\begin{figure}
\caption[Figure 4]{Forward single neutron spectra for Au + Au at RHIC.  The
dot-dashed line corresponds to an isotropic neutron distribution, the dashed
line to a $\sin^2(\theta)$ distribution, and the solid line to the
experimentally measured distribution (see text).}
\label{neu}
\end{figure}
\begin{table}
\caption[Table I]{First order $\sigma_{cd}^1$ and unitarity corrected
$\sigma_{cd}$ (in barns) for Au + Au at RHIC as a function of the
cutoff photon energy $\omega_{max}$ (in MeV).  Calculations are presented for
two values of the lower impact parameter cutoff (see text).}
\normalsize
\begin{tabular}{|r|ll|ll|}
& \multicolumn{2}{c|}{$\sigma_{cd}^1$}
&\multicolumn{2}{c|}{$\sigma_{cd}$} \\
$\omega_{max}$&$b_0 = 15  $&$b_0 = 17.32$&$b_0 = 15 $&$b_0 = 17.32$ \\
\tableline
25 & 1.65 & 1.23 & 1.31 & 1.04 \\
103 & 1.99 & 1.49 & 1.55 & 1.24 \\
440 & 3.10 & 2.32 & 2.29 & 1.84 \\
2000 & 4.21 & 3.16 & 2.98 & 2.42 \\
17840 & 5.12 & 3.84 & 3.50 & 2.86\\
$\infty$ & 5.91 & 4.39 & 3.90 & 3.19 \\
\end{tabular}
\label{tabi}
\end{table}   
\begin{table}
\caption[Table II]{As in Table I, but for Pb + Pb at LHC.}
\normalsize
\begin{tabular}{|r|ll|ll|}
& \multicolumn{2}{c|}{$\sigma_{cd}^1$}
&\multicolumn{2}{c|}{$\sigma_{cd}$} \\
$\omega_{max}$&$b_0 = 15 $&$b_0 = 17.32$&$b_0 = 15 $&$b_0 = 17.32$ \\
\tableline
25 & 2.21 & 1.66 & 1.71 & 1.36 \\
103 & 2.66 & 2.00 & 2.01 & 1.61 \\
440 & 4.12 & 3.09 & 2.92 & 2.37 \\
2000 & 5.58 & 4.19 & 3.76 & 3.09 \\
17840 & 6.79 & 5.09 & 4.42 & 3.65 \\
$\infty$ & 12.75 & 9.45 & 7.15 & 6.02 \\
\end{tabular}
\label{tabii}
\end{table}   
\begin{table}
\caption[Table III]{Parameter dependence of correlated dissociation cross
sections for Au + Au at RHIC.  Units are fm and barns.}
\normalsize
\begin{tabular}{|rllllll|}
$\sigma_{NN}$&$b_0$&R&a&$\sigma_N$&$\sigma_C$&$\sigma_{NC}$ \\
\tableline
50 & 10. & 6.38 & .535 & 6.92 & 6.03 & 10.90 \\
50 & 12.76 & 6.38 & .535 & 6.92 & 4.77 & 10.90 \\
50 & 15. & 6.38 & .535 & 6.92 & 3.92 & 10.74 \\
50 & 17.32 & 6.38 & .535 & 6.92 & 3.17 & 10.09 \\
50 & 12.76 & 6.5 & .535 & 7.09 & 4.77 & 11.01 \\
40 & 12.76 & 6.5 & .535 & 6.97 & 4.77 & 10.93 \\
\end{tabular}
\label{tabiii}
\end{table}   
\begin{table}
\caption[Table IV]{Dependence various cross sections (in barns) on the radius
of the nuclear density $R$.}
\normalsize
\begin{tabular}{|r|ll|ll|}
& \multicolumn{2}{c|}{Au + Au at RHIC}
&\multicolumn{2}{c|}{Pb + Pb at LHC} \\
$R$&$6.38$&$6.50$&$6.624$&$6.75$ \\
\tableline
$\sigma_{1n,1n}$&$.449$&$.445$&$.537$&$.533$ \\
$\sigma_{1n,xn}$&$1.364$&$1.349$&$1.897$&$1.881$ \\
$\sigma_{xn,xn}$&$10.90$&$11.01$&$14.75$&$14.85$ \\
$\sigma_{1n}$&$49.20$&$49.18$&$105.93$&$105.91$ \\
$\sigma_{xn}$&$102.35$&$102.42$&$227.28$&$227.34$ \\
\end{tabular}
\label{tabiv}
\end{table}   
\begin{table}
\caption[Table V]{Single and mutual Coulomb dissociation scaling.  Cross
sections for Ca + Ca and O + O were simply scaled appropriately by $Z^3$ or
$Z^6$ from the calculations for Au + Au at RHIC.  Units are barns.}
\normalsize
\begin{tabular}{|r|ll|ll|}
& \multicolumn{2}{c|}{Single}
&\multicolumn{2}{c|}{Mutual} \\
&All&One Neutron&All&One Neutron \\
\tableline
Au + Au & 95. & 49. & 3.9 & .45 \\
Ca + Ca & 1.5 & .8 & .001 &.0001 \\
O + O & .1 & .05 & .000004 & .0000005 \\
\end{tabular}
\label{tabv}
\end{table}   

\begin{references}
\bibitem {brw} A. J. Baltz, M. J. Rhoades-Brown, and J. Weneser,
Phys. Rev. A {\bf54}, 4233 (1996).
\bibitem {bw} A. J. Baltz and S. N. White, RHIC/DET Note 20, BNL-67127 (1996).
\bibitem {bw1} A. J. Baltz and S. N. White, Summary of the 2nd Zero Degree
Workshop at BNL, February 21-22, 1997.
\bibitem {zero} Proposal to build Forward Calorimeters for the RHIC Heavy Ion
Experiments, www.rhic.bnl.gov/html/zdc.html, August, 1997.
\bibitem {ves} A. Veyssi\`ere, H. Beil, R. Berg\`ere, P. Carlos,
and A. Lepr\^etre, Nuc. Phys. {\bf A159}, 561 (1970).
\bibitem {RHIC} RHIC Design Manual, BNL-52195 (1989).
\bibitem {Alice} Alice Technical Proposal, CERN/LHCC 95-71.
\bibitem {NA49} T. Alber et al. (NA49 collaboration), manuscript in
preparation.
\bibitem {dev} H. deVries, C. W. deJager, and C. deVries, Atomic and Nuclear
Data Tables {\bf 36}, 495 (1987).
\bibitem {gal} C. J. Batty, E. Friedman, and A. Gal, Progress of Theoretical
Physics Supplemant {\bf 117}, 227 (1994).
\bibitem {pdata} R. M. Barnett, et al., Particle Data Group,
Phys. Rev. D {\bf54}, 1 (1996).
\bibitem {tg} Franca Tagliabue and J. Goldemberg, Nuc. Phys. {\bf 23},
144 (1961).
\bibitem {lep} A. Lepr\^etre, H. Beil, R. Berg\`ere, P. Carlos,
J. Fagot, A. de Miniac, and A. Veyssi\`ere, Nuc. Phys. {\bf A367}, 237 (1981).
\bibitem {car} P. Carlos, H. Beil, R. Berg\`ere, J. Fagot, A. Lepr\^etre,
A. de Miniac, and A. Veyssi\`ere, Nuc. Phys. {\bf A431}, 573 (1984).
\end{references}
\end{document}